\documentclass[preprint,prl,aps,a4paper]{revtex4-1}
\usepackage{epsfig}

\newcommand{\red}{\relax}

\begin{document}

\title{Electric activity at magnetic moment fragmentation in spin ice}

\author{D.~I.~Khomskii}

\affiliation{II. Physikalisches Institut, Universit\"at zu K\"oln,
Z\"ulpicher Stra\ss e 77, D-50937 K\"oln, Germany}

\begin{abstract}
Spin ice systems display a variety of very nontrivial properties, the
most striking being the existence in them of magnetic
monopoles.
Such monopole states can also have nontrivial
electric properties: there exist electric dipoles attached to each
monopole.
A novel
situation is encountered in the moment fragmentation (MF) state,
in which monopoles and antimonopoles
are perfectly ordered, whereas spins themselves remain
disordered. We show that such partial ordering strongly modifies
the electric activity of such systems: the electric dipoles, which are
usually random and dynamic, become paired in the MF state in $({\bf d}, -{\bf d})$
pairs, thus strongly reducing their electric activity. The electric
currents existing in systems with noncoplanar spins are also strongly
influenced by MF\null.  We also consider modifications in  dipole and
current patterns in magnetic textures (domain walls, local defects)
and at excitations with nontrivial dynamics in a MF state, which show
very rich
behaviour
and which could in principle allow to control them by electric field.

\end{abstract}

\maketitle

Close connection between electric and magnetic phenomena is a
cornerstone of modern physics, going back to the work of James Clerk Maxwell,
and, even earlier on, to that of Michael Faraday. Recently it acquired
novel significance in the big field of spintronics, which involves such
phenomena as multiferroics, topological systems,  Rashba
spin--orbit coupling,  etc.~\cite{Zuti,Khomskii-Trends,Hasan}.   The magnetic properties of
systems with localised electrons are very diverse. There are
different types of magnetic ordering --- ferromagnetism,
antiferromagnetism, spiral structures, and, similarly, different types of
spin liquids, with their eventual topological features. A very
interesting phenomenon was discovered in spin ice systems such as
some kagome and pyrochlore materials~{\red \cite{Gardner}} --- the appearance of states
resembling the famous magnetic monopoles~\cite{Castelnovo}.
This discovery led to a flurry of activity, both
theoretical and experimental~\cite{Morris,Fennel,Mengotti}.

\begin{figure}
\includegraphics{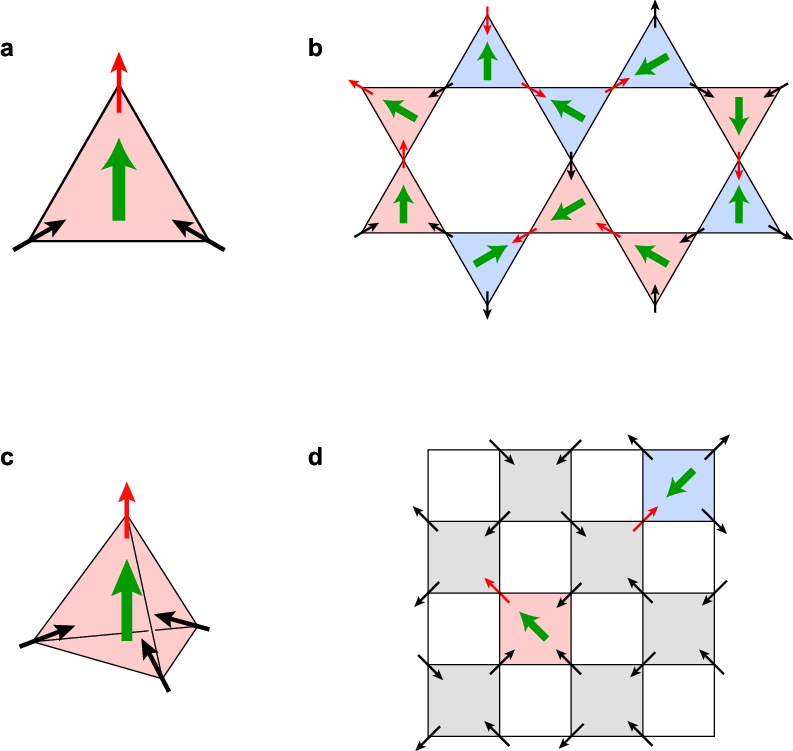}
\caption{\label{FIG:1}{\bf The appearance of an electric dipole on a
magnetic monopole.}  ({\bf a})~The triangle (a building block of kagome ice). The dipole points towards
the ``special''
spin (red arrow).
({\bf b})~Spin ice kagome lattice
with random monopoles ($\mu$) and antimonopoles ($\bar\mu$) and random dipoles.
({\bf c})~The tetrahedron (the building block of pyrochlores) with a monopole.
The dipole  points to the ``special''
spin (red arrow): the out-spin in the monopole with the (3-in)--(1-out) configuration,
and the in-spin in the antimonopole with the (1-in)--(3-out) configuration.
({\bf d})~A typical configuration of
pyrochlore spin ice with random monopoles ($\mu$) and antimonopoles
($\bar\mu$) and the electric dipoles associated with them
shown on a schematic representation of this state on a 2d plot.
The monopoles are marked by pink and antimonopoles by blue colour;
electric dipoles are shown by thick green arrows.}
\end{figure}

It was later realised that for magnetic monopoles in spin ice
there is an interesting interplay between magnetic and electric
degrees of freedom: it was shown that on each magnetic monopole there
exists an electric dipole attached to it~\cite{Khomskii-ice},
see Fig.~\ref{FIG:1}.
This prediction was later
{\red followed up by very interesting theoretical development~\cite{Jaubert,Lantagne-Hurtubise,Slobinsky}, and}
confirmed by different
experimental means~\cite{Grams,Jin}. Thus the interplay
between electricity and magnetism was also demonstrated for magnetic
monopoles.

Recently an interesting new twist in the monopole story appeared: it
was shown that there occurs in these systems a novel state in which
magnetic monopoles exist not only as excitations, but can be also
present in the ground state.  Especially interesting is the
state in which magnetic monopoles and antimonopoles exist on each
metal tetrahedra in pyrochlores  and are ordered in a two-sublattice
fashion (the tetrahedra in pyrochlores and at B-sites of spinels form a
bipartite diamond lattice).
At the same time the spins
themselves remain disordered, see Fig.~\ref{FIG:2} below.

This novel state with such partial
ordering was shown to be a consequence of {\it magnetic moment fragmentation}
(MF)~\cite{Brooks-Bartlett,ElsaLhotel}
(or spin fragmentation state --- not to be confused with spin fractionalization): by the Helmholtz
decomposition the magnetization can be divided into divergence-free and
divergence-full components, and the monopoles and antimonopoles, i.e.\
magnetic charges, the sources of the divergent-full part of magnetisation,
are ordered~\cite{footnote:1}.

This partially-ordered state has a very clear
manifestation in magnetic neutron scattering:
the Bragg scattering due to ordered monopoles (actually forming the (all-in)--(all-out) state) coexists with strong diffuse
scattering with pinch-points due to disordered spins~\cite{Petit,Canals}.
This state can exist in pyrochlores
for a certain range of parameters and temperatures~\cite{Raban};
it can be stabilised by the staggered field present in
specific situations~\cite{Lefrancois,Cathelin}.

A similar state with the ordering of monopoles--antimonopoles
exists also in kagome systems, Fig.~\ref{FIG:2}(b); actually this state was predicted for
kagome systems~\cite{Moller,Chern} even before the idea
of moment fragmentation in pyrochlores was proposed.
In kagome spin ice systems with moment fragmentation the monopoles and antimonopoles
are ordered into two sublattices in a bipartite honeycomb lattice of
triangles.
{\red The kagome ice state can be also realized in pyrochlores such as
Dy$_2$Ti$_2$O$_7$ and Ho$_2$Ti$_2$O$_7$ in a [111] magnetic field slightly below the
transition to the fully-ordered state~\cite{Aoki},
and  it is rarely noticed that this state has moment fragmentation with monopole ordering in kagome
layers, see Supplementary Information~1.}

Spin-ice states in kagome systems are intrinsically the states with
monopoles and antimonopoles in the ground state: every triangle has
either (2-in)---(1-out) spins (monopole, $\mu$, with magnetic
charge $Q=+1$), or (1-in)---(2-out) (antimonopole, $\bar\mu$, $Q=-1$).
That is why the conditions of getting
moment fragmentation in kagome systems is easier to achieve
than in
pyrochlores, in which one first has to create monopoles and antimonopoles, and after that to
order them. Nevertheless the resulting
properties of  these systems are in many respects similar, and in
what follows we will first consider our effects --- the electric
properties of moment fragmentation state --- on the example of kagome
spin-ice systems, and turn to pyrochlores later.

The question we want to address is what happens with charge degrees
of freedom of monopoles, such as electric dipoles, and also with electric
currents and orbital moments, when the system goes to the moment fragmentation state with
ordered monopoles and antimonopoles but with disordered spins.
{\red  The second question we want to consider is the electric
activity of excitations and defects, e.g.\ domain walls (DW) in a
moment fragmentation state, which turns out to be quite nontrivial.}

\section{Results}

We will first quickly recapitulate the arguments leading to the
appearance of electric dipoles ${\bf d}$ attached to magnetic monopoles. As
was shown in \cite{BBMK, Khomskii-chirality}, proceeding from the Hubbard model, in a
magnetic texture  of triangles with spins ${\bf S}_1$, ${\bf S}_2$ and ${\bf S}_3$ there
appears spontaneous charge redistribution (even in a simple strong
Mott insulator with one electron per site and with $U/t\gg1$), so that
the charges at different sites are not equal to $e=1$, but, for example, the
charge at site~1 is
\begin{equation}
e_1 \sim  1 + b [{\bf S}_1\cdot ({\bf S}_2 + {\bf S}_3) - 2 {\bf S}_2 \cdot {\bf S}_3]\;,
\label{eq:e1}
\end{equation}
where the coefficient $b$ in the Hubbard model is given by $b=8t^3/U^2$.
Consequently, if this spin correlation function is nonzero, there
would appear in such a triangle  an electric dipole moment.

\begin{figure}
\includegraphics{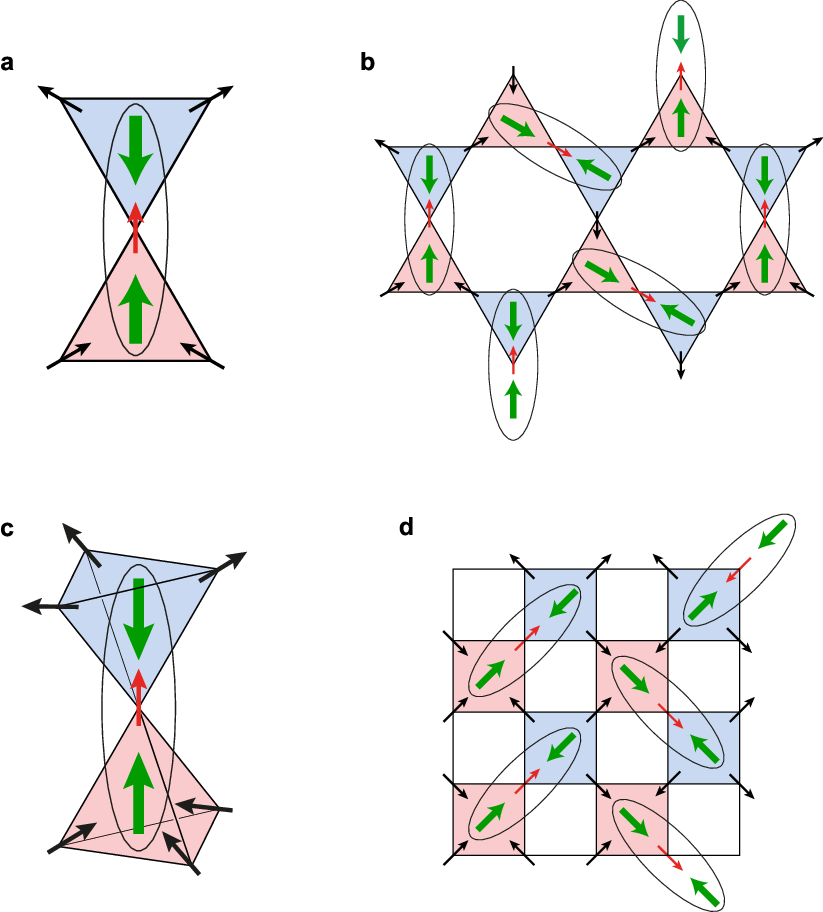}
\caption{\label{FIG:2}{\bf Moment fragmentation state and electric dipoles in this state in kagome and pyrochlore spin ice.}
({\bf a})~Monopole--antimonopole pair (the main building blocks of kagome system with moment fragmentation) with common
``special'' spin (red arrow), demonstrating the pairing of opposite
dipoles (formation of $({\bf d},-{\bf d})$ pairs).
({\bf b})~Example of a moment fragmentation state in kagome spin ice with ordered
monopoles, disordered spins and paired dipoles.
({\bf c})~Pair of tetrahedra with
monopole--antimonopole with a common ``special'' spin, demonstrating the formation of paired $({\bf d},-{\bf d})$ dipoles.
({\bf d})~An example of
moment fragmentation state in a pyrochlore system with ordered
monopoles,
disordered spins and paired dipoles.
The $({\bf d},-{\bf d})$ pairs are marked by ovals.
Colour coding is the same as in Fig.~\ref{FIG:1}.}
\end{figure}

One can easily see that the monopoles and antimonopoles in triangles of
kagome ice always have such dipoles at every triangle, with the dipole moment pointing
in the direction of the ``special'' spin --- the out-spin in the (2-in)--(1-out) monopole,
and the in-spin in the antimonopole with the (1-in)--(2-out)  spin configuration. For pyrochlore systems the (2-in)--(2-out)
states of a tetrahedron (as well as the
(all-in)--(all-out)
states) do not have such dipoles, whereas they are present in monopole and
antimonopole configurations, Fig.~\ref{FIG:2}. The dipole would in this case also point
in the direction of the ``special'' spin (out-spin in the (3-in)--(1-out) monopole
configuration, in-spin in the antimonopole).

In
addition to electric dipoles, always accompanying magnetic monopoles~\cite{Khomskii-ice},
we will also
consider electric currents, which  are present in systems with magnetic
triangles
when the magnetic structure is noncoplanar. As was shown
in \cite{BBMK,Khomskii-chirality}, in this case there exist in such a triangle with spins
${\bf S}_1$, ${\bf S}_2$
and ${\bf S}_3$ a real circular electric current
\begin{equation}
j_{123} = c\kappa(123)
\label{eq:j123}
\end{equation}
where $\kappa$ is the scalar spin chirality,
\begin{equation}
\kappa(123) = {\bf S}_1 \cdot [{\bf S}_2 \times {\bf S}_3]
\label{eq:kappa123}
\end{equation}
and the coefficient $c$ in the nondegenerate Hubbard model is
given by $c=24et^3/\hbar U^2$~\cite{BBMK}. In most spin ice systems the spins are intrinsically
noncoplanar, thus one can expect the appearance of such currents in
triangles in some kagome systems and in tetrahedra in pyrochlores. The
question is thus what would become of these currents (and
the corresponding orbital moments ${\bf L} \sim {\bf j}$),
and of course with electric
dipoles, when the system goes into the moment fragmentation state~\cite{footnote:2}.
As the effects connected with these
currents and with the respective orbital moments have less apparent experimental consequences,
we will present in the main text only the corresponding results,
the detailed  treatment being deferred to Supplementary Information.

{\bf Kagome systems.}    In kagome spin ice, with monopoles and
antimonopoles at every triangle (disordered and mobile in the usual
case)  the electric dipoles exist already in the ground state.
But in the usual situation, without MF, they are random and fluctuating, Fig.~\ref{FIG:1}(b).

This however strongly changes if we go to the moment fragmentation state
with ordered monopoles and antimonopoles. In this case  each triangle
with the monopole $\mu$ is surrounded by three  triangles with antimonopoles $\bar\mu$.
One immediately sees that in this
case there would  always exist monopole--antimonopole ($\mu$--$\bar\mu$)
pairs with the common spin being this ``special'' spin for
both triangles, Fig.~\ref{FIG:2}(a). But as explained above, in this case the electric
dipoles in these triangles would always point towards each other and
would be ``paired'' into a state $({\bf d}, -{\bf d})$ without net dipole moment (but with
nonzero quadrupole moment), see Fig.~\ref{FIG:2}(a).

If however we consider the
$\mu$--$\bar\mu$ pair with different ``special'' spins, the dipole moments
on these triangles would point in different directions. But every
such triangle would then have another ``mate'', another triangle having
this common special spin, having an opposite dipole, so that in effect
in this case {\it all} dipoles would  be paired in $({\bf d}, -{\bf d})$ pairs,
Fig.~\ref{FIG:2}(b). This is actually the main effect we want to stress: the
monopole--antimonopole ordering occurring in a fragmentation state
leads to {\it pairing} of electric dipoles, so that, in particular, the dielectric response
due to these dipoles would be strongly suppressed in the fragmentation
state.

In the typical fragmentation state the spins themselves are
disordered, which means that such $({\bf d}, -{\bf d})$ pairs would be random. Each
such pair connects two neighbouring triangles, i.e.\ it lives on the
bond of the honeycomb lattice of triangles, and every triangle participates in one such bond.  Mathematically this
problem could thus be mapped to the problem of dimer covering of a
honeycomb lattice --- the lattice of triangles in kagome systems. A lot is
known about this problem --- the entropy of this state, etc.~\cite{Elser,Moessner}.
If we go to the long-range
ordered state in which the spins are also ordered, this would
simultaneously give an ordering of these $({\bf d}, -{\bf d})$ pairs, i.e.\ an
ordering of such dimers, or of the corresponding electric quadrupoles.

{\bf Pyrochlores.} \ One can show that the main effects discussed
above would also exist in  pyrochlore spin ice systems with moment
fragmentation.  In this case the fragmentation state with
monopole ordering would also always contain pairs of
($\mu$, $\bar\mu$) with the common ``special'' spin (the spin-out for
monopole with the (3-in)--(1-out) state and the spin-in for the
antimonopole), and correspondingly the dipole moments, pointing to
this ``special'' spin, would again form $({\bf d}, -{\bf d})$ pairs, Fig.~\ref{FIG:2}(c).
Consequently, similar to the kagome case, in a fragmentation
state of the bulk system {\it all} tetrahedra would always be paired,
with  the  electric dipoles at such pair of tetrahedra  pointing
opposite to each other, i.e.\ they would form local $({\bf d}, -{\bf d})$
pairs, Fig.~\ref{FIG:2}(d). Again, these $({\bf d}, -{\bf d})$ pairs would be
random in a fragmentation state with disordered spins, and would be
ordered for ordered spins. Mathematically the problem here would be
the problem of dimer covering of a diamond lattice (lattice of
tetrahedra in pyrochlores).  Experimentally one should see these
effects e.g. in the reduction of electric activity at the  transition
of spin ice pyrochlores to  a MF state.


{\bf Currents and orbital moments in a moment fragmentation state.} \ The
situation with spontaneous currents and corresponding orbital
moments, which may appear at certain magnetic configurations
according to Equations\ (\ref{eq:j123}),~(\ref{eq:kappa123}),
is considered in details in the Supplementary Information.
In short, there appear such currents and orbital moments on monopoles
and antimonopoles in pyrochlore spin-ice systems, and also in
``kagome-from-pyrochlores'' (or tripod kagome), created by the replacement of triangular
[111] layers in pyrochlores by nonmagnetic ions, keeping the kagome [111]
layers magnetic~\cite{Paddison}. Spin orientation in such systems follows the same
rule as in the underlying pyrochlores, i.e.\ the spins in both systems
are noncoplanar, which leads to nontrivial effects connected with
currents and orbital moments. The treatment presented in the Supplementary Information shows
that the transition to a moment fragmentation state also modifies
the pattern of currents and respective orbital moments, in such a way that these
orbital moments ${\bf L}$ are {\it perfectly ordered} in kagome-from-pyrochlore
systems,
and they are paired (with parallel $({\bf L}, {\bf L})$ moments) in pyrochlores,
but with such pairs remaining {\it disordered}. In
general, orbital moments are parallel (or antiparallel) to the net
spin of respective triangles or tetrahedra, i.e.\ at first glance
they would not lead to any noteworthy
physical effects, 
but there may be interesting effects connected with the scalar spin chirality, see below.


{\bf Defects  and excitations in a moment fragmentation state.} \ Yet
another interesting and important aspect to
consider is what would be the electric properties of excitations,
defects and textures in the moment fragmentation with ordered ($\mu$, $\bar\mu$)
configurations. As always, when there is an ordered state of some kind,
one can form different types of defects: local defects, similar to an
inverted spin in an antiferromagnet, or domain walls between
domains, which differ by the interchange of ($\mu$, $\bar\mu$) sublattices. In
addition, in contrast to the usual two-sublattice systems such as N\'eel
antiferromagnets, here there may appear yet other types of local
excitations: in the kagome MF states with (2-in)--(1-out) configurations,
there may appear (3-in) or (3-out) states of a triangle, which would be
a novel type of defect.  Similarly, such states can appear in
pyrochlores: besides ordered ($\mu$, $\bar\mu$) configurations in the MF
state, there may exist defects with (4-in) or (4-out) tetrahedra, or the simple
spin-ice states (2-in)--(2-out).

Usually such defects are formed in pairs by reversing one spin.
By that we exchange magnetic charges at neighbouring sites (by sites we mean
triangles in kagome and tetrahedra in pyrochlore systems). Thus when we reverse
a ``special'' spin in the structures of Fig.~\ref{FIG:2}(a),~(c), we create a pair
of ``supermonopoles'' --- a pair of (all-in)--(all-out)
defects, triple
monopoles (``tri-poles'') in kagome systems of Fig.~\ref{FIG:2}(a), with
charges $+3Q$ and $-3Q$: and (4-in) and (4-out) states (double monopoles) in
pyrochlores, Fig.~\ref{FIG:2}(c), with charges $+4Q$ and $-4Q$. (In pyrochlores
it is common to define the monopole
charge~$q$ as $\frac12$ of that given above, so that the monopole would have
charge $q$ $(=+1)$, and the (4-in) state (double monopole) would have
charge~$2q$). When we invert not a ``special''  but the usual spin in a
MF state, we create in kagome a pair of defects with interchanged monopoles
and antimonopoles, i.e.\ with monopoles at the ``wrong'' sublattice.
In pyrochlores we create by that two usual spin ice states, with (2-in)--(2-out)
tetrahedra, with charges~0. In both cases the reversal of one spin changes
the charges at adjacent sites (triangles, tetrahedra) by $\pm2Q$, the total
charge being of course conserved. Thus one can say that for example the charge
of a tri-pole in kagome case is~$3Q$, but also that {\it the excess charge}
is~${2Q}$, as compared with the charge~$Q$ of the original monopole at this
site.  Such excitation, when it moves in a crystal,
carries with it the excess charge~${2Q}$. This language is very convenient for
considering the motion of such excitations in the MF state.

Once created, these defects can move in a crystal, similar to the case
of the usual monopoles in spin ice.
The motion of these excitations in a MF state is rather nontrivial.
As this question, though definitely interesting,  lies somewhat aside
from the main topic of this paper --- the electric activity of such defects
and excitations ---  I schematically present this discussion in the Supplementary
Information; see also~\cite{Jaubert2}.
What is important is that such defects can move in a MF state without
confinement, similar to the motion of monopoles in the usual spin ice~\cite{Castelnovo},
although with certain restrictions.
Therefore we can consider the electric activity of isolated defects of this type.

\begin{figure}
\includegraphics{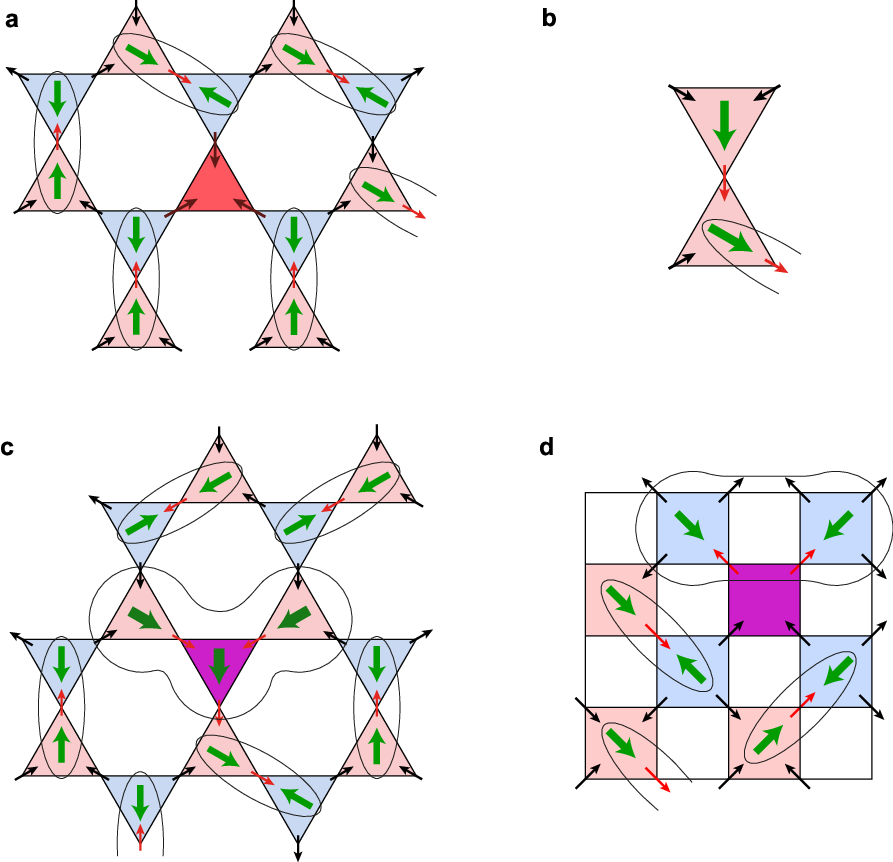}
\caption{\label{FIG:3}
{\bf Dipole structure at defects and monopole textures in kagome moment fragmentation state.}
({\bf a})~Triangle with (3-in) state (``tri-pole''). Reconnection of free spins allows to create such defect without free dipoles.
({\bf b})~Typical situation with the monopole at the ``wrong'' place, at the
antimonopole sublattice, demonstrating the appearance of unpaired dipoles.
Note that the common spin is the ``normal'' in-spin for
one triangle, but it is a ``special'' out-spin for the other triangle.
Consequently the dipole
moment at the second, upper triangle would point to this ``special'' spin and would
be definitely unpaired.
({\bf c})~Monopole in place of an antimonopole (magenta triangle). One sees that
it inevitably leads to the creation of three unpaired dipoles:
on the defect triangle itself and on two (out of three) of its neighbours.
({\bf d})~The appearance of (two) unpaired dipoles next to (2-in)--(2-out) tetrahedron
(magenta) in pyrochlore with moment fragmentation.
}
\end{figure}

Consider first kagome systems, and start from a defect of the
type (3-in) with the charge~$3Q$ in place of a monopole (2-in)--(1-out),
or (3-out), charge~$-3Q$,
instead of an antimonopole. This situation is shown in Fig.~\ref{FIG:3}(a).

As mentioned above, such ``tri-poles'' (monopoles with charge~$3Q$,
or with the excess charge~${2Q}$, can move in their own sublattice by
interchanging some spins, and by moving they leave a trail of changed spins,
but such strings have no tension, see Supplementary Information.
As to their eventual dipole activity, first one easily
sees using Equation~(\ref{eq:e1}) that there would be no charge redistribution
and no electric dipoles at such ``tri-pole'' triangles themselves.
This is also evident just from the symmetry (all three sites and all
directions in such a triangle are equivalent). In principle the
creation of such a dipole-less triangle (or a defect site in the
effective honeycomb lattice of triangles) would remove one dipole
from a $({\bf d}, -{\bf d})$ pair, so that one could think that there would
appear one
unpaired dipole.  However, by reversing some spins one can in this
case ``recommute'' the remaining ($\mu,\bar\mu$) pairs so that all
dipoles would again be paired.
And actually one can see that when such tri-pole is one of a pair
created at neighbouring sites (triangles) by reversing a special spin
(the tri-pole that moved away from its partner), this pairing of dipoles in between
those occurs automatically during the motion of such defects, see Supplementary Information~1.
And the string connecting these defects is thus ``decorated'' by the reversed
$({\bf d}, -{\bf d})$ pairs.

\begin{figure}
\includegraphics{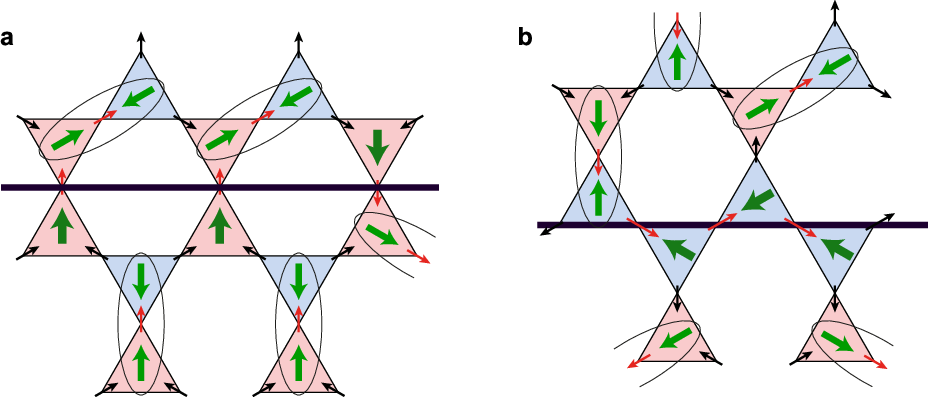}
\caption{\label{FIG:4}
{\bf Dipole structure of domain walls in kagome systems with moment fragmentation.}
({\bf a}),~({\bf b})~Monopole domain walls of type 1 and~2.
There are unpaired dipoles on domain walls of both types.
The situation would be in principle similar for pyrochlore systems with moment fragmentation
(shown in Fig.~S4 of the Supplementary Information).}
\end{figure}

The situation might be different for other defects or textures, e.g.\ created
by reversing the usual, not a ``special'' spin. The
main building block of those is the neighbouring pair of triangles
both with monopoles (or antimonopoles). As one sees in Fig.~\ref{FIG:3}(b), in this
case a common spin is the in-spin for one triangle, but it is the
out-spin (i.e.\ our red ``special'' spin) for the other, upper triangle. In effect in
this upper triangle the electric dipole would point to this special
spin and would not have a ``mate'', an opposite dipole at
a neighbouring ``site'', i.e.\ it would not form a $({\bf d}, -{\bf d})$
pair --- it would definitely be an unpaired dipole.

Using this rule it is easy to understand the electric properties of different
defects and monopole textures like domain walls. The
simplest of those could be the interchange of $\mu$, $\bar\mu$ at
neighbouring sites, or just the  reversal of a monopole charge at
some site, e.g.\ the replacement of $\bar\mu$ by $\mu$ at a particular
triangle. The typical resulting situation is shown in Fig.~\ref{FIG:3}(c).
(One can easily show that such defect can also move in
its sublattice without leading to confinement --- there would be no
string tension on its trajectory either, see Supplementary Information~1.) As to the electric activity of such
reversed monopole, one sees that in this case one unavoidably creates
unpaired dipoles. A typical building block in this case is formed by three
neighbouring triangles all with monopoles (or all with
antimonopoles). It is clear that for all possible spin configurations
one necessarily has some unpaired dipoles, see Fig.~\ref{FIG:3}(c). In
effect in a kagome lattice such defects would lead to the creation of even three
unpaired dipoles, one at the central ``reversed'' triangle, and two dipoles at two out
of three adjacent triangles. Thus such defects ($\mu$ at a $\bar\mu$
position, or vice versa) would necessarily carry unpaired dipoles.

Actually the situation is very similar on domain walls created by
the interchange of $\mu$ and $\bar\mu$ in a part of the ordered MF state.
There are two types of such domain walls, shown in Fig.~\ref{FIG:4}(a) and~(b).
At a domain wall of the first type, Fig.~\ref{FIG:4}(a), we have a pair of
($\mu,\mu$) or ($\bar\mu, \bar\mu$) triangles of the same type, as
in Fig.~\ref{FIG:3}(b). As is explained there, in this case we
necessarily have an  unpaired dipole moment perpendicular to such
domain wall, Fig.~\ref{FIG:4}(a). These dipoles can be random, up or down.  The
situation is similar for the domain wall of the second type,
Fig.~\ref{FIG:4}(b), except that such neighbouring ($\mu, \mu$) triangles would be
oriented differently --- more or less along the domain wall (in a tilted zigzag
pattern). Consequently also in this case, by the same reason, there
would appear unpaired dipoles along
such domain wall, Fig.~\ref{FIG:4}(b).

In fact the situation is rather similar for defects and spin
textures in pyrochlores with moment fragmentation. By the same reason ---
partial ordering (ordering of monopoles with still free spins),
local defects are, first of all, mobile, i.e.\ it does not cost
a linearly increasing energy to move them in their own sublattice ---
due to the freedom to reverse spins on the MF state
there would be no confinement in this case. Defects of the (4-in) (or
(4-out)) type  would again have features of monopoles with a larger
magnetic charge: monopoles with (3-in)--(1-out) state have magnetic
charge~${+2Q}$, and the (4-in) state has charge~$+4Q$,  i.e.\ in this case the
{\it excess charge} of the excitation itself is also~$2Q$.

As to dipole activity, one can again see that defects of this
type --- (4-in) tetrahedra at a $\mu$-site or (4-out) at a $\bar\mu$ site of
the MF state should not necessarily create any dipoles: such sites (tetrahedra)
themselves are symmetric and do not have any dipoles; and one can always recommute
spins on neighbouring tetrahedra to get rid of eventual unpaired
dipoles.  And again, when we first create a pair of such double monopoles
with charges $+4Q$ and~$-4Q$ by reversing the special spin in Fig.~\ref{FIG:2}(c)
(effectively destroying the dipoles of the $({\bf d},-{\bf d})$ dimer on these
neighbouring sites), then moving these defects apart we automatically recommute
spins in such a way as to create on the trajectory of these defects
new $({\bf d},-{\bf d})$ pairs, see the Supplementary Information.

On the other hand,  the ``non-monopole'' spin ice (2-in)--(2-out) defects,
created e.g.\ by reversing the usual spins
(also mobile without confinement) would
again, similar to the kagome case, create an unpaired dipole (actually two
of them), see Fig.~\ref{FIG:3}(d).
As is clear from this figure, two neighbouring tetrahedra adjacent to the
(2-in)--(2-out) defect (which in itself has no electric dipole) should
necessarily have unpaired dipoles directed towards such defect.

For other types of defects or textures, e.g.\ for domain walls, the main
``building block'' is two neighbouring tetrahedra
with one common site both having monopoles (or both antimonopoles).
Similar to the situation of Fig.~\ref{FIG:3}(b), this pair also necessarily has at least
one unpaired dipole directed towards the ``special'' spin of the common site, the dipole at
neighbouring tetrahedron pointing away from it, so that at least the
first dipole cannot have a ``mate'' with opposite direction.
Therefore the situation in monopole domain walls for pyrochlores is also
similar to that in kagome systems. In this case
there would also be present nn pairs of tetrahedra both with
monopoles (of both with antimonopoles), and by the same reason as
above there should necessarily exist unpaired dipoles at such domain
walls. The details of their orientation are described in the Supplementary Information.
The interchange of spins inside every domain can lead to
reversal of some such dipoles, so that generically the monopole domain
walls in the MF state should have extra electric activity, both in kagome
and in pyrochlore spin ice systems with MF.


\section{Discussion}
In this paper we demonstrated that in spin ice systems (kagome, pyrochlores) there exist quite nontrivial
electric properties when
there occurs in them a moment fragmentation --- the formation of a partially
ordered state in which monopoles and antimonopoles are present on
every triangle in kagome and every tetrahedron in pyrochlores and are
ordered in a two-sublattice fashion, while spins themselves are still
disordered. This partial ordering strongly influences electric
dipoles attached to every monopole and antimonopole in both these systems.
Whereas without moment fragmentation these dipoles are random and dynamic
(fluctuating together with spins), in the fragmentation state they are
paired into $({\bf d}, -{\bf d})$ pairs on neighbouring triangles or tetrahedra.
Therefore the transition to a fragmentation state should lead to a
reduction of electric activity present in spin ice systems without
fragmentation due to random and dynamic dipoles, which is seen e.g.\
in microwave absorption~\cite{Grams}.


On the other hand, electric
currents existing in triangles with noncoplanar spins~\cite{BBMK,Khomskii-chirality}
behave differently in kagome and in pyrochlore systems. In kagome
systems (``kagome-from-pyrochlores'', see Supplementary Information)
such currents and the orbital moments associated
with them, random in the usual case, become
long-range ordered simultaneously with the ordering of monopoles at
moment fragmentation. In pyrochlores, however, such currents and the
corresponding orbital moments in the fragmentation state remain random,
but are also paired, but in this case  into $({\bf L}, {\bf L})$ pairs.

The formation of staggered current pattern and staggered scalar spin
chirality in ``kagome-from-pyrochlores''  may lead to quite
nontrivial effects. It is known that the existence of scalar spin
chirality and the corresponding Berry phase leads to the appearance of
very strong fictitious magnetic field. For total nonzero spin
chirality it leads, in particular, to the intrinsic anomalous Hall
effect~\cite{Taguchi}.


Very interesting and at first glance surprising is the electric
behaviour of different defects in a MF state. Such point defects or
excitations, like the ``supermonopoles'' ((all-in) and (all-out)
states)
can in principle be generated by external disorder, impurities etc., but
they can also be excited thermally with the excitation energy different
from system to system but typically of the order of the transition/crossover
temperature to the MF state.  Thus e.g.\ in Dy$_2$Ir$_2$O$_7$
this excitation energy is~${\sim}\,3.6\,\rm K$, whereas the crossover to the MF
state occurs at~${\sim}\,1.4\,\rm K$~\cite{Cathelin}.
Probably one can also create such excitations by external perturbations, e.g.\ by
microwave irradiation.

In
contrast to naive expectations not all such defects break $({\bf d}, -{\bf d})$
pairs and lead to a creation of unpaired dipoles. The defects of monopole
type with larger magnetic charge (``tri-poles'' in kagome --- (3-in) or (3-out)
triangles) and ``tetrapoles'' in pyrochlores ((4-in) or (4-out)
tetrahedra) do not in general lead to unpaired dipoles. However other
defects or textures --- e.g.\ (2-in)--(2-out) states in pyrochlores and monopoles on ``wrong'' antimonopole sites ---
breaks the $({\bf d}, -{\bf d})$ pairs and  necessarily lead to the creation of
unpaired dipoles. The same is true for domain walls in the $(\mu,\bar\mu)$
ordered state in both kagome and pyrochlore systems: these
also carry unpaired dipoles. Thus the creation of such defects or
textures generically should be accompanied by an increase of
electric activity of moment fragmentation systems.


Such defects in
the partially-ordered fragmentation state have their own interesting
dynamics --- typically they can be moved without confinement, similar
to monopoles in the usual spin ice.
And at high concentration such defects would start to interact with each
other, in particular by the dipole-dipole interaction, which can lead to rather nontrivial behavior~\cite{Jaubert}.

The study of defects and their
dynamics, both magnetic and electric, in the moment fragmentation state is
yet experimentally a completely unexplored field which definitely
deserves the interest of specialists.

{\red Let us now consider some possible experimental
manifestations of the effects discussed above.

1.~The main effect is the pairing of electric dipoles in a MF state
into $({\bf d},-{\bf d})$ pairs, both in pyrochlore and in kagome spin ice. The
most apparent consequence is the corresponding reduction of electric
activity in this state, e.g.\ measured by microwave absorption, as
used e.g.\ in~\cite{Grams}. In pyrochlores the most reliable observation of a
MF state was made in Ir pyrochlores such as Dy$_2$IrO$_7$, in which
the (all-in)--(all-out) ordering of Ir at high temperatures gives
staggered field for monopole sublattices~\cite{Lefrancois}, so that
in this case the transition to a MF is not a sharp transition but
rather  a crossover with broad anomalies~\cite{Cathelin}. In
principle the effects could be more prominent in systems in which
the MF state would appear as a real well-defined transition. However
e.g.\ the results claiming such transition in  Nd$_2$Zr$_2$O$_7$~\cite{Petit} were
recently questioned~\cite{Opherden}, and it seems that such transition
without external field is difficult to realise in pyrochlores. The
situation in this sense may be better in kagome ice systems. There
are reliable reports that such transition was indeed observed in
these systems, both natural and artificial~\cite{Canals,Paddison}. The existence of
such well-defined MF phase ``sandwiched'' between the disordered
high-temperature phase and the fully ordered phase at low temperatures
was also obtained theoretically~\cite{Moller,Chern,Canals}. Kagome systems may be
preferable also from another point of view. In pyrochlores the monopoles are the excited states,
thus to observe the effects discussed above (such as the suppression of
electric activity when going to the MF state) one first has to create
monopoles with attached electric
dipoles, e.g.\ by changing the temperature, magnetic field etc., which
would increase the electric activity, and then to order the monopoles in a MF
state, forming $({\bf d}, -{\bf d})$ pairs, which would lead to a reduction of this
activity. In kagome ice systems, on the other hand, monopoles with
their dipoles exist in all phases, and the reduction of electric
activity in going to a MF state would be seen more clearly.

The
formation of dipoles at monopoles in spin ice is associated with
the corresponding shifts of oxygens sitting in the middle of metal
tetrahedra; one sees these shifts e.g.\ by Raman scattering~\cite{Jin}.
The transition to a MF state would then lead, together with the
formation of $({\bf d}, -{\bf d})$ pairs, to correlation of these shifts at
adjacent tetrahedra or triangles. These correlated distortions could
probably be also investigated by Raman scattering or by some other
technique.

2.~But probably even more interesting would be the experimental
situation for monopole textures and defects, such as monopole
domain walls.  The unique feature of a MF state is the coexistence in
one spin subsystem of an ordered and a disordered components. As always,
each time we have long-range ordering in a system, the most
interesting effects are connected with excitations violating this
order~\cite{Khomskii-Book}. However just for a MF state, with its remaining spin
disorder, such DW would be rather elusive, there would be no
prominent magnetic contrast on them. Probably just because of that
there was no study, even no mention of such DW in the literature.  And
the fact that, as shown above, such DW (and also many point
defects) would  be ``decorated'' by electric dipoles, would make
their observation possible, for example by STM, sensitive to local
electric fields thus created (e.g.\ using electron standing waves~\cite{Becker}). And the electric
activity of such DW could in principle allow one to influence and control
them by electric field, as was done e.g.\ for ordinary DW in
ferromagnets in~\cite{Logginov}. Similarly one could think of controlling point
defects with uncompensated dipoles by an inhomogeneous electric field,
cf.\ the study of skyrmions by such method in~\cite{Hsu}.

3.~As is shown above, in particular in SI, the MF state should also
display nontrivial patterns of scalar spin chirality and of
spontaneous currents and respective orbital moments. Somewhat unluckily,
in all cases considered the direction of these orbital moments $\bf L$ for
every tetrahedron or triangle in pyrochlore and kagome systems turned
out to be parallel to the respective net spin of these,
${\bf S}_{\rm tot}$, and typically the orbital moments would make only a
small fraction of the full moment ${\bf M} = {\bf S}_{\rm tot} + {\bf L}$.
Thus, I don't expect
interesting purely magnetic effects connected with those. But on the
other hand we get here a nonzero scalar spin chirality~$\kappa$, Eq.~(\ref{eq:kappa123}),
the same for every triangle in kagome-from-pyrochlore systems, see
Fig.~S6.
Similarly, such ``ferro'' scalar chirality would exist in
kagome ice state, created in spin ice pyrochlores in the [111] magnetic
field slightly below the transition to the fully-ordered state, see e.g.~\cite{Aoki}.
And one should expect that this scalar spin chirality would
give the intrinsic (Berry phase) anomalous Hall effect, similar to
the case of Nd$_2$Mo$_2$O$_7$~\cite{Taguchi}.  This effect would be very interesting to
check experimentally.}

In conclusion we want to stress once again a very special and unusual
feature of moment fragmentation state in spin ice: the coexistence
{\it in one spin system} of one component (monopoles, a
divergence-full part of magnetization) with perfect long range
ordering, and a disordered component (divergence-free part of
magnetization). This strange situation with partial ordering leads to
many unusual properties, such as the existence of novel types of defects and
excitations, domain walls with special properties, etc.  And we
demonstrated that the electric activity of this state is also very
special: on one hand, due to ordering in the monopole--antimonopole
sector the electric dipoles are paired into $({\bf d}, -{\bf d}$ pairs,
and, on the other hand, due to spin disorder these dipole pairs are
random and dynamic. The same dichotomy of the moment fragmentation
state is also reflected in the electric properties of defects and
domain walls. This two-faced Janus
character of this state makes these systems especially interesting and unusual.

\newpage

\renewcommand{\thefigure}{S\arabic{figure}}
\setcounter{figure}{0}

\renewcommand{\vec}[1]{{\bf #1}}

\section{Supplementary information}

\begin{figure}
\includegraphics{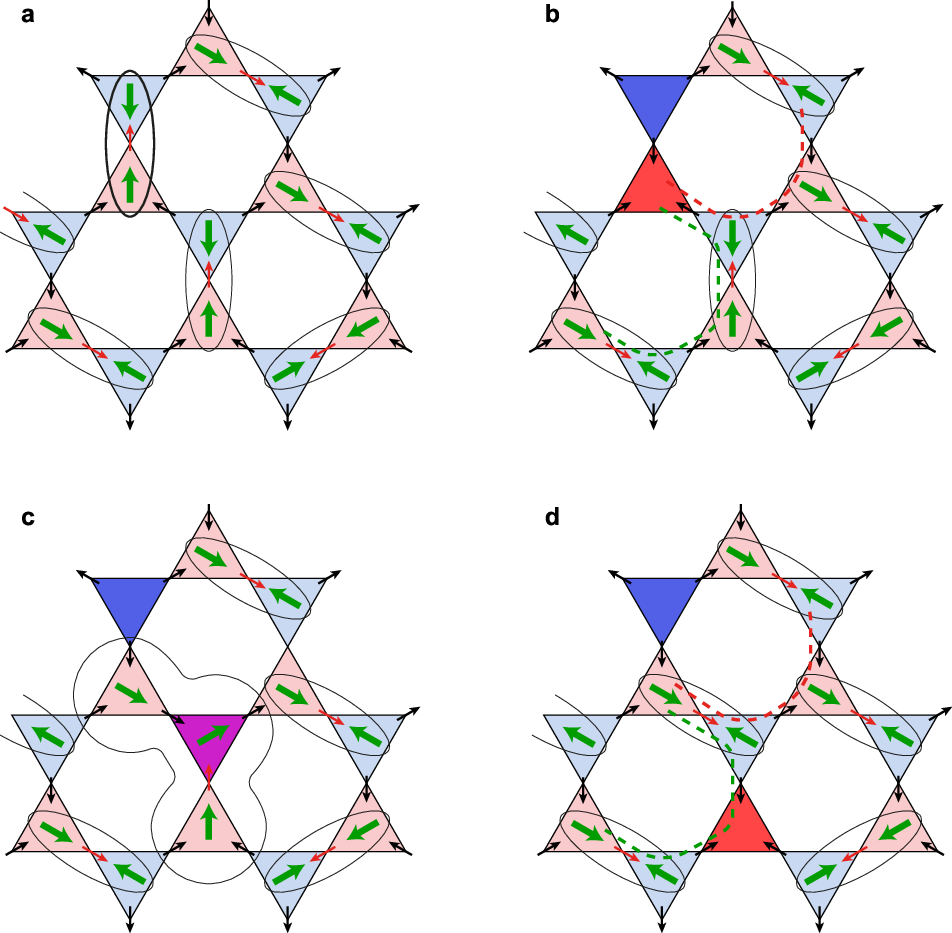}
\caption{\label{FIG:S1}
{\red Creation and motion of excitations in kagome systems with moment fragmentation and the corresponding
distribution of electric dipoles.
({\bf a})~Initial state with ordered monopoles, with dipoles (thick green arrows) paired in $(\vec d,-\vec d)$ dimers (ovals).
({\bf b})~Creation of triple monopoles and antimonopoles (red and blue triangles) by reversing one ``special'' spin
(red arrow), with ``annihilation'' of two dipoles at the respective triangles.
({\bf c}),~({\bf d})~Motion of an excitation with the excess charge~$+2Q$ by consecutive reversal of one spin at each step.
In ({\bf c}) it leads to the creation of a monopole (magenta triangle) in an antimonopole sublattice,
with the creation of three unpaired dipoles, cf.~Fig.~3(c). In ({\bf d}) another ``special'' spin is reversed,
and in effect the tri-pole (red triangle) shifts in its own sublattice leaving a trail of reversed spins
(and paired dipoles) without
confinement and without unpaired dipoles. By the green dashed line we show an allowed trajectory of the excitation,
a string without confinement; by the red dashed line we show a forbidden trajectory, on which
wrong sites (triangles) would appear.}}
\end{figure}

{\red
{\bf 1. Creation and motion of defects in systems with moment fragmentation and their dipole character.}

Point defects and the corresponding excitations in MF states can be
created by some perturbations, e.g.\ by external defects etc., but they can
also be thermally excited. In the later case they are created in
pairs by reversing some spins.  As discussed in the main text if one
reverses the ``special'' spin say in Figs.~2(a) and~2(c), one creates
a pair of (all-in) and (all-out) sites (triangles in kagome,
tetrahedra in pyrochlore systems). The charges of these objects are,
respectively, $\pm3Q$ and $\pm4Q$.   Note right away that by making this
pair of defects we simultaneously ``destroy'' two electric dipoles
which originally formed a $(\vec d, -\vec d)$ dimer pair on these sites. Another
possibility is to reverse a ``normal'', not a special spin. In kagome
systems this will interchange the monopole and antimonopole, i.e.\
create a ($\bar\mu, \mu$) pair at the wrong places, in wrong sublattices.
In pyrochlore systems such reversal of the usual spin will create
two tetrahedra of the (2-in)--(2-out) type. As discussed in the main
text, in both cases the defects created by that have some unpaired
dipoles, though for example in pyrochlores the (2-in)--(2-out) sites
themselves have no dipoles, see Fig.~3(c),~(d).

By making such spin reversal we either create ``supermonopoles'' ((3-in)
states in kagome, (4-in) states in pyrochlores), with the
increased charge, or we instead decrease the charges at sites, e.g.\ by
making non-monopole (2-in)--(2-out) tetrahedra in pyrochlores.   One
can speak in this case not of the charge of the defect itself, but
of the {\it excess charge} of an excitation: one gets e.g.\ the ``double
monopole'' of (4-in) type with charge~$4Q$ instead of the original state
of a monopole with charge~$2Q$, i.e.\ the excess charge of such
excitation is~$+2Q$. This language is very convenient when
considering the motion of such excitations.

Once created, these defects can start to move in a crystal by
consecutively reversing some spins one after the other.  But, in
contrast to a similar motion of monopoles in regular spin ice~[5], here
these excitations move on the background of ordered monopoles. This
makes their motion more complicated and more interesting.

If the motion of such defects would occur via simultaneous flipping
of two spins~[33], the defects can remain in their own sublattice.                     
But the usual motion occurs via flipping of one spin at a time. As
shown in Fig.~\ref{FIG:S1}, in this case at each step the excitations moves from
one sublattice to the other and by that always change their character
(including their dipole structure).  But they carry their excess
charge with them. Thus, the excitation having the form of a tri-pole
of Fig.~\ref{FIG:S1}(b) with charge~$3Q$ and excess charge~$+2Q$ moves to
the neighbouring site, Fig.~\ref{FIG:S1}(c), and at this site its charge is
$-Q + (+2Q) = +Q$, i.e.\ instead of the original antimonopole at this
site (the blue triangle in Fig.~\ref{FIG:S1}(a),~(b)) it becomes a monopole with
charge~$+Q$ (the magenta triangle in Fig.~\ref{FIG:S1}(c)).  Simultaneously, as is
seen form this figure, one creates the defect with three unpaired
dipoles, as in Fig.~3(c) of the main text.
}

\begin{figure}
\includegraphics{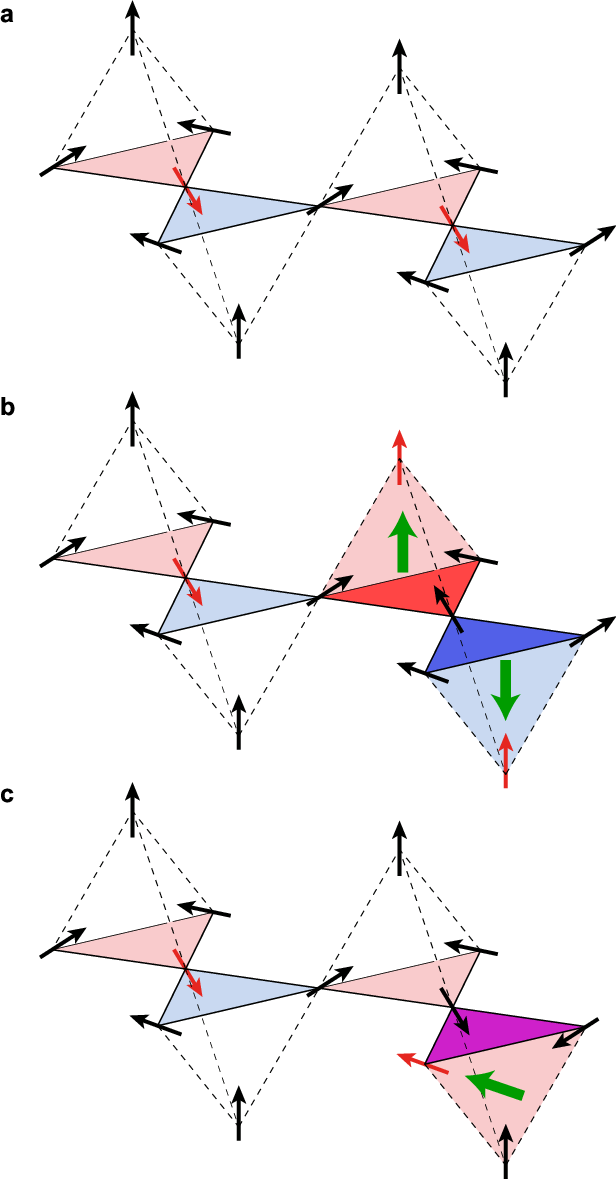}
\caption{\label{FIG:S2}
{\red Ground state and defects in kagome ice, obtained in pyrochlore systems such as Dy$_2$Ti$_2$O$_7$ in magnetic
field $\vec H \parallel [111]$ slightly below the transition to a fully-ordered state.
({\bf a})~Picture demonstrating that such a state is actually a MF state with ordered monopoles in the kagome plane.
({\bf b}),~({\bf c})~Two types of defects in the kagome plane: ({\bf b})~tri-poles, ({\bf c})~monopole in the wrong sublattice (magenta triangle).
The electric dipoles at these defects are shown by thick green arrows.}}
\end{figure}

{\red
After that the defect has two options. Either it can move forward by
reversing again the ``special'' spin, Fig.~\ref{FIG:S1}(d), or by reversing the
remaining usual spin. In the first case one creates the tri-pole at a
site of a monopole sublattice, i.e.\ one effectively moves the
tri-pole from one site of this sublattice, red triangle in Fig.~\ref{FIG:S1}(b), to another
site of the same sublattice,
Fig.~\ref{FIG:S1}(d). This motion occurs via the intermediate ``virtual'' state
when this excitation is on the ``wrong'' sublattice, Fig.~\ref{FIG:S1}(c), but in effect
one can move in this way the excitation in its own sublattice. And
one sees that by such motion we change the spins, and also electric
dipoles, on the trajectory of the defect, but, similar to the case of
the usual spin ice, all monopole states on such trajectory return to
their original state, i.e.\ such string has no tension and there is no
confinement. However if on the second step one would reverse not the
special spin, as is done in going from (c) to~(d), but the remaining
usual spin of the magenta triangle in Fig.~\ref{FIG:S1}(c),
moving the defect ``to the right'' in Fig.~\ref{FIG:S1}(c), then one
would leave a trail of wrong monopoles on such trajectory. Thus, in
contrast to monopoles in the usual spin ice, which can move without
confinement in all directions, here the motion is partially
restricted: at every second step the defect can only move in one
direction, in the direction of the respective special spin.
In effect not all trajectories of excitations are allowed, see Fig.~\ref{FIG:S1}.
And this
guarantees two things simultaneously: the absence of confinement, and
also the absence of a trail of unpaired dipoles: all electric dipoles
on the trajectory of the defect would be also paired in $(\vec d, -\vec d)$ pairs
--- except maybe the ends of these strings, which would have unpaired
dipoles if they are of the type of monopole on a wrong place in
kagome and (2-in)--(2-out) defect in pyrochlores; these would have
unpaired dipoles and would consequently contribute to the electric
activity of the system.

Thus, in effect, when such point defects move on the MF background,
they change the character all the time, in kagome changing sublattices, and in pyrochlores from the
double monopole to (2-in)--(2-out) states. I.e.\ this is actually the
motion of the excitation with the charge~$+2Q$ (or~$-2Q$) on the
periodic potential: such defects have different energies on different
sublattices. Which state has lower and which has higher energy may depend
on the particular situation. Thus one may think that in pyrochlores the
(2-in)--(2-out) state would have lower energy than the (4-in) state. But
the situation may be different for example in a MF state which exists in the kagome ice
state in pyrochlores in [111] magnetic field slightly below phase
transition to the fully-ordered state [25], see Fig.~\ref{FIG:S2}: the (3-in)
(tri-pole) state of a kagome triangle (which is simultaneously a
monopole (3-in)--(1-out) or (1-in)--(3-out) state of a tetrahedron)
would have lower energy than the state of kagome monopoles in place
of antimonopoles. In the first case all spins in kagome planes (their
$z$-components) would point along the field, whereas for the second
case two out of three such spins would be opposite to it. Thus in this case
one should expect that the energy of an excitation
would have lower energy when it is a ``tri-pole''.
(Note, by the way, that such defects in kagome ice in a [111] field in
pyrochlores would also have electric dipoles, shown in Fig.~\ref{FIG:S2},
whereas such kagome ice state itself has here no dipoles
(every tetrahedron in this case has the (two-in)--(two-out) configuration without dipoles).)
All in all,
this is a very interesting situation deserving special attention
(see also {\red [33]}), which, however, lies beyond the scope of the
present paper.
}

{\bf 2. Defects in pyrochlores with moment fragmentation.}

\begin{figure}
\includegraphics{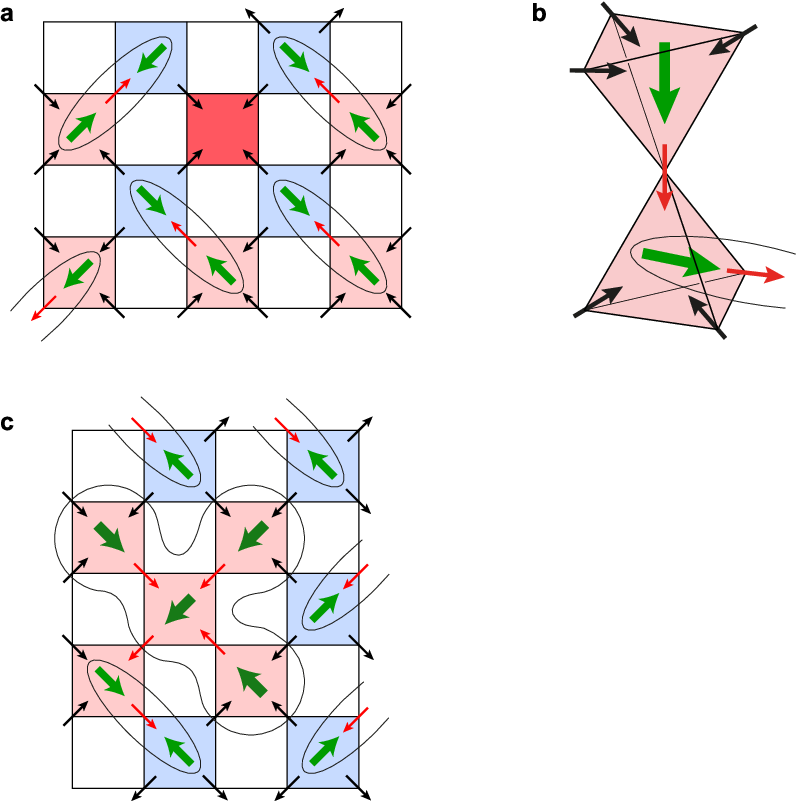}
\caption{\label{FIG:S3}
Dipoles and defects in pyrochlore spin ice.
({\bf a})~``Supermonopole'' (4-in state).
({\bf b})~Typical situation with two monopole neighbours. The common spin is the ``special''
out-spin for one monopole, and the dipole pointing in this direction is unpaired.
({\bf c})~Point defect --- the monopole in place of an antimonopole. Four unpaired dipoles would form at such defect.}
\end{figure}

Similarly to the case of kagome systems discussed in the main
text, one can see that in pyrochlores the defects have in principle
very similar properties, as to their dipole activity.   In Fig.~\ref{FIG:S3}(a) we show
the structure of the first type of point defects: the ``supermonopole'' state
of the type (4-in) or (4-out). One sees from this figure that by
recommuting the remaining spins one can form such defect without any
free unpaired dipoles. But this is not the case
for the other types of  defect. {\red One of them, the (2-in)--(2-out) state, shown
in Fig.~3(d) of the main text, has, as is shown there, two unpaired dipoles
attached to it. These two defects, (4-in) and (2-in)--(2-out) states, are the
most important point defects, which, when they move, interchange all the time,
see the previous section~SI~1.}

For other types of defects or textures, e.g.\ for domain walls, very important
is the situation with two neighbouring tetrahedra both of monopole or
antimonopole type. As one sees from Fig.~\ref{FIG:S3}(b),
in this case a common spin is the in-spin for one tetrahedron, but it is the
out-spin (i.e.\ our ``special'' spin) for the other tetrahedron. In this second tetrahedron
the electric dipole would point to this special spin, and it would not form a $(\vec d, -\vec d)$
pair --- it would definitely be an unpaired dipole.

Using this picture it is easy to understand that e.g.\ the isolated point
defect --- monopole in place of an antimonopole, would lead to the formation
of even four such unpaired dipoles, Fig.~\ref{FIG:S3}(c): this ``wrong''
monopole tetrahedron would have four other neighbouring tetrahedra
also with monopoles, and, by the arguments presented above, see
Fig.~\ref{FIG:S3}(c), four out of these 5 tetrahedra would have unpaired spins.
{\red Such defect, which would have an excess charge~$\pm4Q$, is of high energy,
it requires the flipping of two spins, thus as an isolated defect it probably
does not play an important role (but it can be created e.g.\ by the ``fusion''
of two usual excitations with the excess charge~$2Q$).  But such states play
an important role in domain walls.}

\begin{figure}
\includegraphics{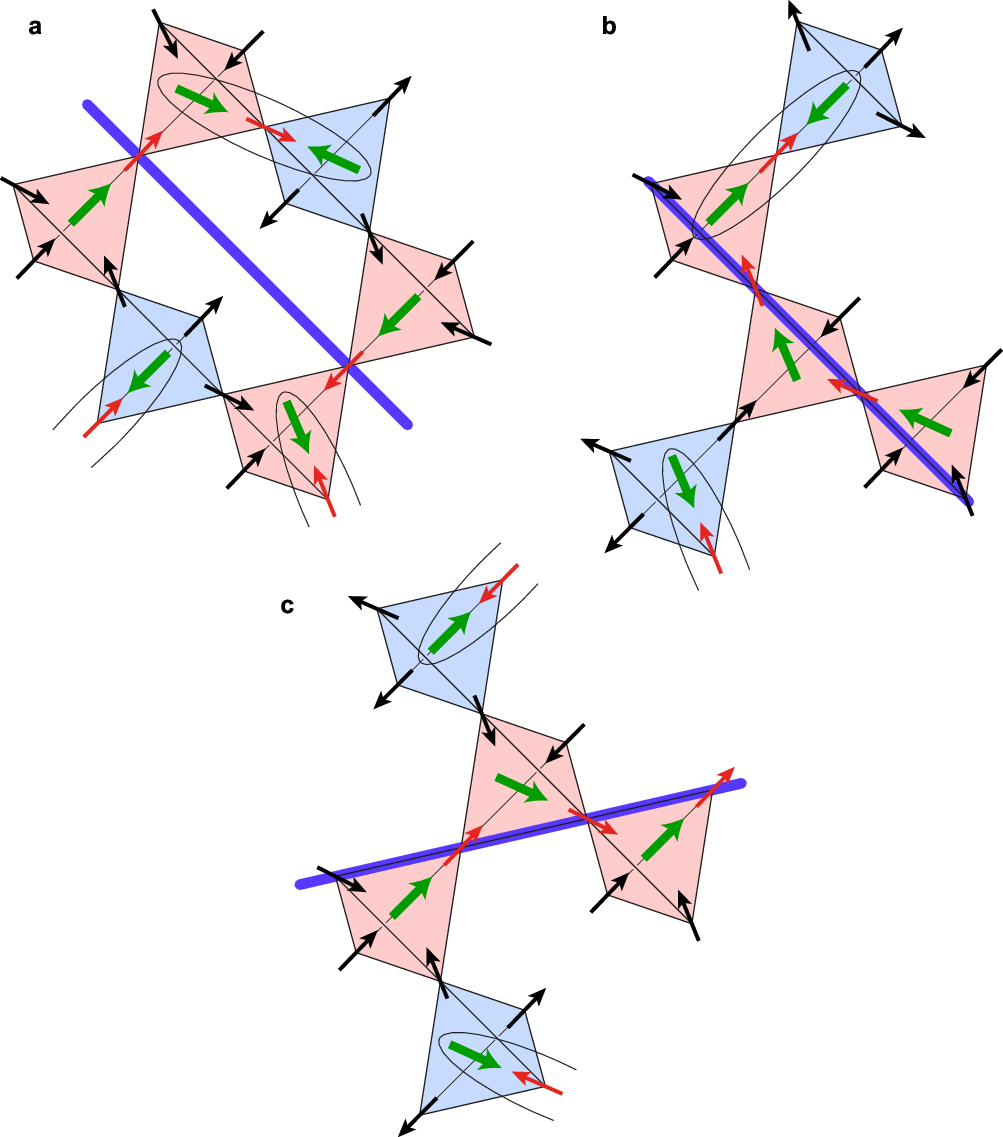}
\caption{\label{FIG:S4}
Domain walls in pyrochlores with moment fragmentation, with corresponding electric dipoles.
({\bf a})~Domain wall running through triangular (111) plane.
({\bf b})~Domain wall running through kagome (111) plane.
({\bf c})~Domain wall at (001) plane.}
\end{figure}

The same arguments show that there would appear unpaired dipoles at
domain walls in pyrochlores with moment fragmentation. There would be
in this case three types of such domain walls due to interchange of
$\mu$ and $\bar\mu$ sublattices in half of the sample, Fig.~\ref{FIG:S4}. Domain
walls of the first type, similar to domain walls of type~1 in kagome
systems, are the walls going through triangular sites in (111) layers
in pyrochlore lattice, Fig.~\ref{FIG:S4}(a). At such walls the neighbouring
tetrahedra would have one common site, and dipoles would point
perpendicular to such domain wall, i.e.\ in the [111] direction.

Another domain wall is also perpendicular to the [111] axis, but runs
through the kagome (111) plane, Fig.~\ref{FIG:S4}(b). In this case, similar to the
domain wall of type~2 in kagome systems, we would also have
neighbouring monopole tetrahedra, which again would have
unpaired dipoles, but these dipoles would point more or less along the
domain wall plane (tilted from it). And finally, the third type of
domain walls in pyrochlores with MF, Fig.~\ref{FIG:S4}(c), are those
perpendicular to the cubic [001] axis (or other equivalent cubic
axes). One easily sees that unpaired dipoles in this case would also
point more or less along the plane of a domain wall, as in the previous
case.  Thus, in general, the properties of defects in pyrochlores
with MF are very similar to those in kagome systems, considered in
the main text, and most of the defects would have an electric
activity and could in principle be controlled by an electric field.

{\bf 3. Currents and orbital moments in a moment fragmentation state.}

\begin{figure}
\includegraphics{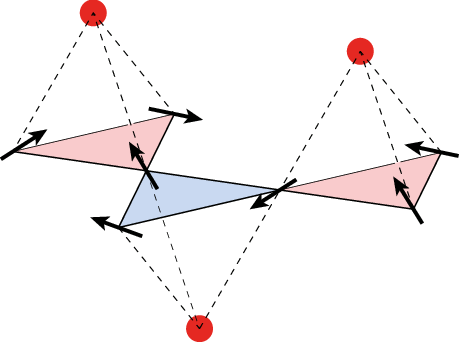}
\caption{\label{FIG:S5}
Structure of a ``kagome-from-pyrochlore'' system,
e.g.\ Dy$_3$SbMg$_2$Sb$_2$O$_{14}$ [31]. Red balls are nonmagnetic ions.}
\end{figure}

As is clear from Eqs.\ (2),~(3), nontrivial effects such as spontaneous
orbital currents and corresponding orbital moments exist only for
magnetic structures with noncoplanar spins. Consequently they are
absent e.g.\ in kagome systems with spins in the $xy$-plane. There exists
however a very interesting group of kagome systems ---
``kagome-from-pyrochlores'' (or ``tripod kagome'') obtained by nonmagnetic dilution of
spin-ice pyrochlores, e.g.\
Dy$_3$SbMg$_2$Sb$_2$O$_{14}$~[31]. One can visualise these
systems as ordered [111] kagome layers of magnetic ions, triangular
[111] layers being occupied by nonmagnetic Sb, see Fig.~\ref{FIG:S5}.

That is, in the original magnetic tetrahedra with Ising ions (spins
pointing in or out of tetrahedra) one makes tetrahedra in which the
apical magnetic ions are replaced by nonmagnetic ones, but the spin
direction of the remaining magnetic ions is the same as before: spins
point either in or out of $M_3M'$ (here Dy$_3$Sb) tetrahedra, see
Fig.~\ref{FIG:S5}.

\begin{figure}
\includegraphics{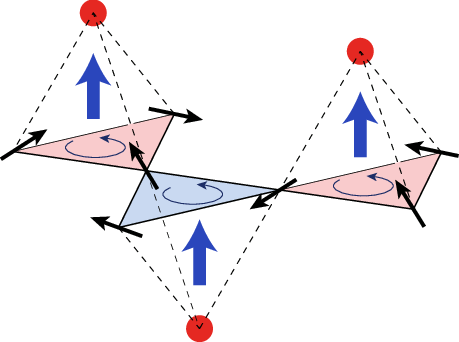}
\caption{\label{FIG:S6}
Orbital currents and orbital magnetic moments in ``kagome-from-pyrochlore'' spin ice in a moment fragmentation state.
Currents (thin rings in triangles), clockwise in monopoles, anticlockwise in antimonopoles (looking from outside of tetrahedra!),
and orbital moments (thick blue arrows) in triangles.
In contrast to electric dipoles, here the
spontaneous currents and orbital moments are fully ordered in a MF state.}
\end{figure}

According to Eqs.\ (2), (3) the scalar spin
chiralities and corresponding currents and orbital moments are of
one sign (e.g.\ currents clockwise, orbital moments up) for monopoles,
and are opposite for antimonopoles. In regular kagome spin ice
(``kagome-from-pyrochlores'') they are random, but in
the moment fragmentation state with full $(\mu, \bar\mu)$ ordering these
currents and orbital moments are also fully ordered, Fig.~\ref{FIG:S6}.  In
that sense currents and orbital moments in MF kagome systems are
different from electric dipoles in those: dipoles form pairs, but are
random and dynamic, but currents and orbital moments are fully
ordered together with monopoles themselves.

\begin{figure}
\includegraphics{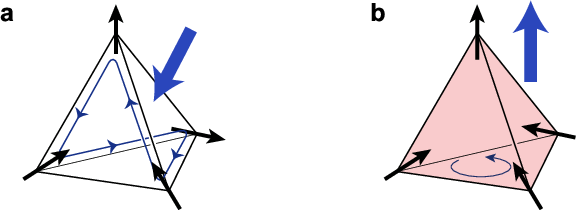}
\caption{\label{FIG:S7}
Orbital currents and orbital magnetic moments in pyrochlore spin ice.
({\bf a})~The usual (2-in)---(2-out) tetrahedra.
({\bf b})~Currents and orbital moments at monopoles and antimonopoles (for the negative coefficient $c$ in Eq.~(2)).
Note that in both cases orbital magnetic moments are parallel (or antiparallel) to the net spin moment of corresponding tetrahedra.}
\end{figure}

The situation with spontaneous currents and orbital moments in
pyrochlores, with and without MF, is a bit more tricky than
that in kagome systems, but conceptually similar. One can show that
for spin ice tetrahedra there would exist nonzero currents even in a
pure spin-ice case (2-in)--(2-out) without monopoles. Indeed, using the
expressions (2), (3), one can show that
the currents in this case would have a trajectory shown in Fig.~\ref{FIG:S7}(a),
i.e.\ there would exist an orbital moment pointing from the edge with
2-in to the edge with 2-out spins. In effect, similar to the kagome
case, the orbital moment would point along (here antiparallel) the
net spin moment of such tetrahedra.

The same rule actually applies to tetrahedra with monopoles or
antimonopoles. The currents in this case would only flow along a
triangle  with (3-in) or (3-out) spins, Fig.~\ref{FIG:S7}(b) (the currents on the
other three edges of such tetrahedra, each belonging to two
triangles, would cancel). And the corresponding orbital moment would
point again in the direction of the total spin of corresponding
tetrahedra.

\begin{figure}
\includegraphics{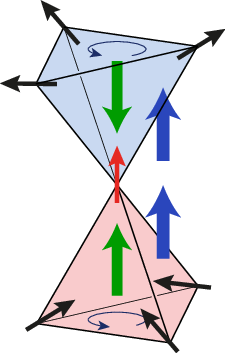}
\caption{\label{FIG:S8}
Currents and orbital magnetic moments for a typical configuration in pyrochlore spin ice with moment fragmentation.
Green arrows are dipole moments, thick blue arrows are orbital moments.}
\end{figure}

The situation with currents and orbital moments in pyrochlores with
moment fragmentation is different from that in a kagome ice.  When we
look at a pair of neighbouring tetrahedra with opposite monopole
charges, Fig.~\ref{FIG:S8}, and with  the ``special'' spin at the common site
between those, we indeed see that the electric dipoles of these
tetrahedra are opposite and form a $(\vec d, -\vec d)$ pair oriented
along the corresponding axis (one of [111] axes of pyrochlore
lattice). But the currents here, according to
Eq.~(3), would run along triangles in the basal
plane, opposite to the ``special'' spin, but they would run (looking
from outside!)\ in the opposite direction, clockwise and
counterclockwise.  Consequently the orbital moments created by these
currents would also be parallel and would point in the same direction
as the total spin of respective tetrahedra.  Thus for monopoles the
orbital moments would be parallel to electric dipoles, $\vec L _i
\parallel \vec d_i$, but for antimonopoles they would point in
opposite directions, Fig.~\ref{FIG:S8}. When spins themselves change (keeping
the MF state), both $(\vec d, -\vec d)$ and $(\vec L, \vec L)$ pairs
move and would change orientation. This is in contrast to the case of
kagome systems (kagome-from-pyrochlore), in which in the
fragmentation state currents and orbital moments are long-range
ordered together with monopoles themselves, whereas the $(\vec
d,-\vec d)$ pairs fluctuate together with spins.

\section*{Acknowledgements}
I am very grateful to L.~Chapon and M.~Mourigal for introducing me
to the field of moment fragmentation in spin ice, and to T.~Michely and G.~Jackeli for useful
discussions. This work was funded by the Deutsche
Forschungsgemeinschaft (DFG, German Research Foundation) -- Project
number 277146847 -- CRC 1238.


\begin{thebibliography}{99}
\bibitem{footnote:1}{\red As is clear from Ref.~\cite{Brooks-Bartlett}, such
decomposition of total magnetization into divergence-free (Coulomb phase) and
divergence-full part is valid even for random monopoles. However in the
follow-up literature the terminology ``magnetic fragmentation'' is always used
for the state with ordered monopoles and antimonopoles at every site (every
tetrahedron in pyrochlores or every triangle in kagome systems). In this paper
we will also use the term MF in this sense.}
\bibitem{footnote:2}Note that one often discusses the fictitious magnetic moment or
magnetic flux, accompanying the states with nonzero scalar spin
chirality due to Berry phase mechanism, see e.g.~\cite{Taguchi}.
But it is interesting and not
usually appreciated, that there appear in this case also real
currents and real orbital magnetic moments proportional to scalar
spin chirality in~\cite{BBMK}.
\end{thebibliography}
\end{document}